\documentclass[aps,pra,twocolumn,preprintnumbers,amsmath,amssymb,footinbib]{revtex4}
\usepackage{mathtools}
\usepackage{lipsum}
\usepackage{mathtools,amssymb,lipsum, nccmath}
\usepackage{physics}

\DeclareMathAlphabet{\mathcalligra}{T1}{calligra}{m}{n}
\DeclareMathAlphabet{\mathpzc}{OT1}{pzc}{m}{it}

\usepackage{graphicx,epsfig}
\usepackage{bm}
\usepackage{dcolumn}
\usepackage{xcolor}% coloured text
\usepackage[breaklinks=true,colorlinks,citecolor=blue,linkcolor=blue,urlcolor=blue]{hyperref}

\def\rmp#1#2#3{Rev. Mod. Phys. {\bf #1}, #2 (#3)}

\def\noi{\noindent}
\def\bc{\begin{center}}
	\def\ec{\end{center}}
\newcommand{\bea}{\begin{equation}}
\newcommand{\eea}{\end{equation}\noi}
\newcommand{\ber}{\begin{eqnarray}}
\newcommand{\eer}{\end{eqnarray}\noi}
\begin{document}
\title{Gauss' law for arbitrarily moving Gaussian surfaces: A revisit}

\author{Shyamal Biswas}\email{sbsp@uohyd.ac.in}
\affiliation{School of Physics, University of Hyderabad, C.R. Rao Road, Gachibowli, Hyderabad-500046, India}
	
\date{\today}
	
\begin{abstract}
We have re-examined the integral form of Gauss’ law for arbitrarily moving charges inside and outside
an arbitrarily expanding (or contracting) and deforming Gaussian surface. We have explicitly calculated the time-dependent Gauss' flux integral for such a generic non-static case with the Maxwell equations under consideration. We have obtained an evolution equation $\frac{\text{d}}{\text{d}\text{t}}\oint_{s(t)}\vec{E}\cdot\text{d}\vec{s}(t)=\frac{I^{(s)}_{\text{in}}(t)}{\epsilon_0}$ for the time-dependence of the flux-integral. We have demonstrated pedagogically that while the flux integral is dependent on the expansion/contraction of the surface, it is independent of its deformation.
\end{abstract}

%\pacs{01.55.+b General physics,  03.50.De Classical electromagnetism, Maxwell equations, 01.40.Fk Research in physics education}
%Keywords: Gauss' law, Gauss' flux-integral theorem, Moving Gaussian surface, Electrodynamics, Maxwell equations, Exact results, Physics education
	
\maketitle	

\section{Introduction}
Classical electrodynamics is a well-established branch of physics and engineering since its one and half centuries old (1865 AD) formulation with the Maxwell equations \cite{Born}. While the integral form of Gauss' law (1835 AD) is a pillar for electrostatics, the differential form of Gauss' law is a pillar for electrodynamics because it is one of the four Maxwell equations \cite{Born}. Classical electrodynamics is a very common undergraduate/postgraduate course for both physics and engineering students. The classical electrodynamics with exotic or semi-classical properties of the optical, dielectric, magnetic, conducting, \textit{etc} media, however, is still a hunting ground for theoretical and experimental research \cite{Garcia-Vidal,Basov,Lapine,Rotter,Garcia-Vidal2}. 

The integral form of Gauss' law in electrodynamics is found in the textbooks \cite{Rothwell2}. However, students often ask questions on the integral form of Gauss' law in electrodynamics especially when the charges inside the Gaussian surface move arbitrarily and become radiative. Questions may also come if the Gaussian surface moves further. These issues are already discussed either implicitly for a general case \cite{Rothwell} or explicitly for some special cases \cite{Garcia}. However, the issues with the separate effect of the expansion/contraction and that of the deformation of the moving Gaussian surface on Gauss' flux integral have not been analysed so far.

In this article, we would like to revisit the integral form of Gauss' law for a generic time-dependent case of charges moving arbitrarily inside and outside a moving Gaussian surface with the Maxwell equations under consideration and analyse the effects of the expansion/contraction and deformation of the Gaussian surface on the time-dependence of Gauss' flux integral, separately.

Let us follow the usual notation in the field such as $\vec{E}=\vec{E}(\vec{r},t)$ ($\vec{B}=\vec{B}(\vec{r},t)$) is the electric field (magnetic field) at the position $\vec{r}$ at time $t$ in the 3-D Euclidean (free) space where $\rho(\vec{r},t)$ and $\vec{J}(\vec{r},t)$ represent the charge density and the current density, respectively. The Maxwell equations for the electromagnetic field in electrodynamics are as follows:
\begin{eqnarray}\label{eqn1}
\div\vec{E}(\vec{r},t)=\frac{\rho(\vec{r},t)}{\epsilon_0},
\end{eqnarray} 
\begin{eqnarray}\label{eqn2}
\curl\vec{E}(\vec{r},t)=-\frac{\partial\vec{B}(\vec{r},t)}{\partial t},
\end{eqnarray}
\begin{eqnarray}\label{eqn3}
\div\vec{B}(\vec{r},t)=0,~\text{and}
\end{eqnarray} 
\begin{eqnarray}\label{eqn4}
\curl\vec{B}(\vec{r},t)=\mu_0\vec{J}(\vec{r},t)+\mu_0\epsilon_0\frac{\partial\vec{E}(\vec{r},t)}{\partial t}.
\end{eqnarray}
Here-from, one can easily show that the electromagnetic wave (light) propagates with the speed $c=1/\sqrt{\mu_0\epsilon_0}\simeq3\times10^8$ m/s in the free space \cite{Born}.

\begin{figure}
\includegraphics[width=8.5cm]{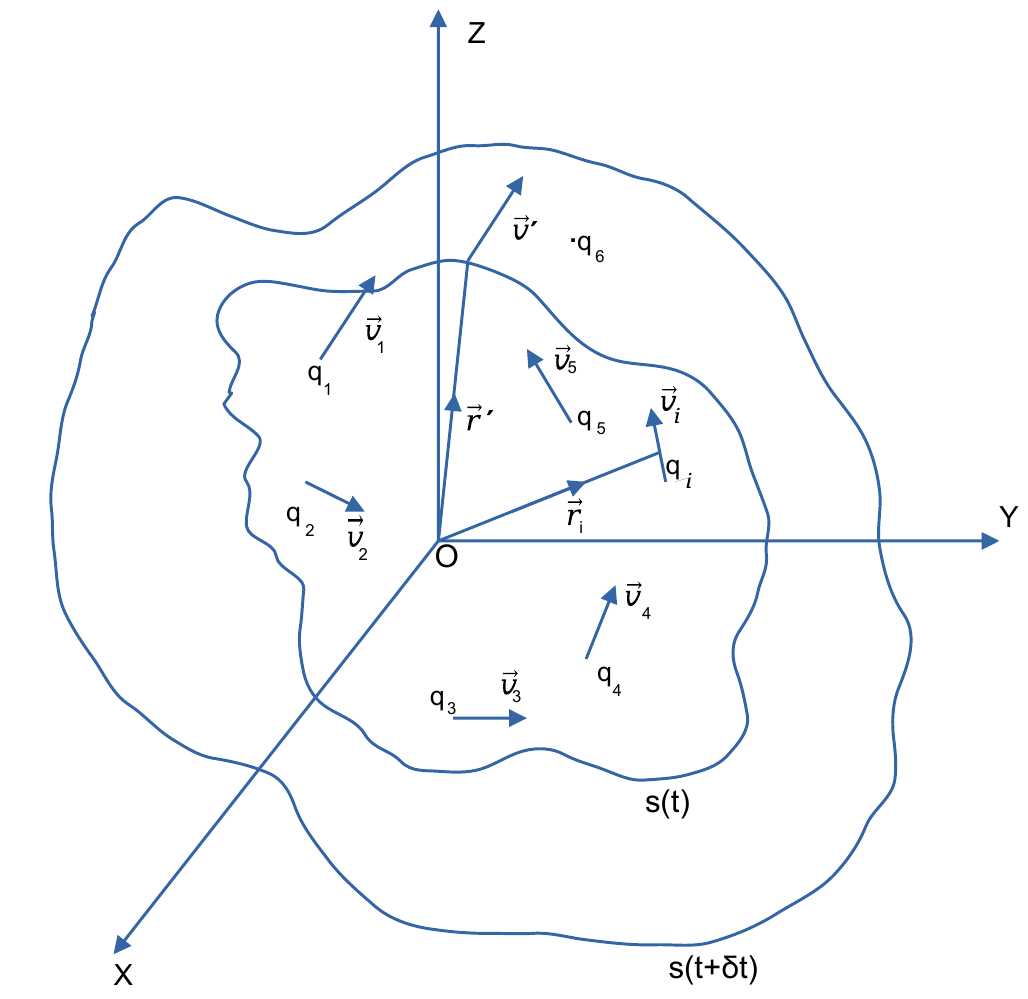}
\caption{A schematic diagram for $i=1,2,3,4,5$, and $6$ with $n=6$. The (point) charges $q_2$, $q_3$, $q_4$, and $q_5$ are moving with velocities $\vec{v}_2$, $\vec{v}_3$, $\vec{v}_4$, and $\vec{v}_5$, respectively, inside the moving and deforming Gaussian surface $s(t)$ at time $t$. The surface becomes $s(t+\delta t)$ at time $t+\delta t$. While the charge $q_1$ is about to leave $s(t)$ with velocity $\vec{v}_1$, the charge $q_6$ which although is at rest, is about to come inside $s(t)$ due to motion of the surface, say, with the velocity $\vec{v}'$ at the point $\vec{r}'$ at time $t$. 
}
\label{fig1}
\end{figure}

While the differential form of Gauss' law is given by Eqn.~(\ref{eqn1}) even in a non-static case  (i.e. $\frac{\partial\rho}{\partial t}\ne0$ and or $\frac{\partial\vec{J}}{\partial t}\ne0$), the integral form of Gauss' law (Gauss' flux theorem or simply Gauss' law) is given by the following form
\begin{eqnarray}\label{eqn5}
\oint_s\vec{E}(\vec{r})\cdot\text{d}\vec{s}=\frac{q_{\text{in}}}{\epsilon_0}
\end{eqnarray}
in the standard textbooks \cite{Jackson,Griffiths} for a static case (i.e. for $\frac{\partial\rho}{\partial t}=0$ and $\frac{\partial\vec{J}}{\partial t}=0$) of the total charge $q_{\text{in}}$ inside a static (closed) Gaussian surface $s$ where a point is denoted by $\vec{r}'$. However, if we integrate Eqn.~(\ref{eqn1}) with respect to $\text{d}^3\vec{r}$ in the volume $V(t)$ and apply Gauss' divergence theorem on the left-hand side, we can easily get
\begin{eqnarray}\label{eqn5a}
\oint_{s(t)}\vec{E}(\vec{r},t)\cdot\text{d}\vec{s}(t)=\frac{q_{\text{in}}(t)}{\epsilon_0}
\end{eqnarray}
where $q_{\text{in}}(t)=\int_{V(t)}\rho(\vec{r},t)\text{d}^3\vec{r}$ represents the total (arbitrarily moving) charge inside the volume $V(t)$ at any instant of time $t$ and $s(t)$ represents the arbitrary moving Gaussian surface which encloses the volume $V(t)$ in the free space as described in figure \ref{fig1}. Eqn.~(\ref{eqn5a}) must be the time-dependent form of Gauss' law which is available in the literature \cite{Rothwell2}. However, the time $t$ is used here as a parameter in both $s(t)$ and $V(t)$ while deriving Eqn.~(\ref{eqn5a}) from Eqn.~(\ref{eqn1}). Such a parametrization neither explicitly takes the motion of the surface $s(t)$ nor explicitly takes the motion of the charges inside and outside $s(t)$ into account. Thus, a doubt remains whether Eqn.~(\ref{eqn5a}) is true for a generic non-static case. Hence, we are motivated to explicitly verify the validity Eqn.~(\ref{eqn5a}) and to provide a transparent derivation of the same by the explicit use of the motion of both the Gaussian surface $s(t)$ and the charges inside and outside $s(t)$.

Laws of electrostatics and magnetostatics such as Coulomb's law and Biot-Savart law, respectively, were already generalized for electrodynamics with the time-dependence under consideration \cite{Griffiths2}. The differential form of Gauss' law, i.e.  Eqn.~(\ref{eqn1}), on the other hand, is valid for a generic case of electrodynamics \cite{Born}. Thus, a natural question comes -- whether the differential form of Gauss' law, i.e. Eqn.~(\ref{eqn5}), is also valid for a generic non-static (dynamic) case of the charges moving arbitrarily inside and outside an arbitrarily moving Gaussian surface. Special cases of such a study were separately discussed (i) for a single charge uniformly moving and (ii) for an accelerated radiating charge, inside a static Gaussian surface under the consideration of the retarded scalar potential \cite{Garcia}.

Our calculation in this article starts with the time-derivative of flux-integral under the consideration of the Maxwell equations as described in Refs.\cite{Jackson1,Zangwill}\footnote{Please note that the time-derivative of flux-integral was described in Refs.\cite{Jackson1,Zangwill} only for the magnetic field passing through an unclosed moving surface.}. Here-from we obtain an evolution equation for the time-dependence of the flux-integral for the electric field in a generic non-static case of both the charges and the Gaussian surface where-in the total amount of the charges is not necessarily a non-conserved quantity. Then we integrate this equation and obtain time-dependent Guass' flux integral. Finally, we conclude.

\section{Gauss' flux-integral in electrodynamics}
In a static case, the left-hand side of Eqn.~(\ref{eqn5}) represents the flux of the electric field passing through the Gaussian surface $s$. On the other hand, in a non-static case as depicted in figure \ref{eqn1}, the flux of the electric field passing through a moving (expanding/contracting and deforming) Gaussian surface $s(t)$ can be given by the time-dependent integral $\oint_{s(t)}\vec{E}(\vec{r},t)\cdot\text{d}\vec{s}(t)$. This flux-integral may vary with time as the surface element $\text{d}\vec{s}(t)$ at the position $\vec{r}'=\vec{r}$ at time $t$ moves with a velocity $\vec{v}'(\vec{r},t)$ and the (point) charges $\{q_i\}$ ($i=0,1,2,3,...,n$) both inside and outside $s(t)$, as depicted in figure \ref{fig1}, move with the velocities $\{\vec{v}_i(\vec{r}_i,t)\}$. By notation, we have $\vec{r}=\vec{r}'$ for $\vec{r}$ on the surface $s(t)$. Otherwise, $\vec{r}$ denotes a point inside and outside $s(t)$. The time-derivative of the flux-integral can be obtained by following the procedure described in Refs.\cite{Jackson1,Zangwill} with the convective derivative ($\frac{D}{D t}=\frac{\partial}{\partial t}+\vec{v}'\cdot\grad$) of a vector field (say $\vec{E}(\vec{r},t)$) for a moving surface as
\begin{eqnarray}\label{eqn6}
\frac{\text{d}}{\text{d}\text{t}}\oint_{s(t)}\vec{E}(\vec{r},t)\cdot\text{d}\vec{s}(t)&=&\oint_{s(t)}\text{d}\vec{s}(t)\cdot\bigg[\frac{\partial\vec{E}(\vec{r},t)}{\partial t}+\vec{v}'(\vec{r},t)(\nonumber\\&&\div\vec{E})-\curl{[\vec{v}'(\vec{r},t)\times\vec{E}(\vec{r},t)]}\bigg].\nonumber\\
\end{eqnarray}
While the 2nd term on the right-hand side of Eqn.~(\ref{eqn6}) is coming from the expansion (or contraction) of the Gaussian surface $s(t)$, the 3rd term on the same side is coming from its deformation \cite{Jackson1,Zangwill}. Now, using the Maxwell equations, especially Eqns.~(\ref{eqn1}) and (\ref{eqn4}), we recast Eqn.~(\ref{eqn6}) as
\begin{eqnarray}\label{eqn7}
\frac{\text{d}}{\text{d}\text{t}}\oint_{s(t)}\vec{E}\cdot\text{d}\vec{s}(t)&=&c^2\oint_{s(t)}\text{d}\vec{s}(t)\cdot\curl\bigg[\vec{B}-\frac{\vec{v}'\times\vec{E}}{c^2}\bigg]\nonumber\\&&+\frac{1}{\epsilon_0}\oint_{s(t)}\text{d}\vec{s}(t)\cdot\big[\rho\vec{v}'-\vec{J}\big].
\end{eqnarray}
It is to be noted that $\vec{J}\neq\rho\vec{v}'$ because $\vec{v}'$ does not represent the velocity of a charge, rather it represents the velocity of the Gaussian surface. Let us now apply Gauss' divergence theorem to the 1st term on the right-hand side of Eqn.~(\ref{eqn7}) to get the following
\begin{eqnarray}\label{eqn8}
\frac{\text{d}}{\text{d}\text{t}}\oint_{s(t)}\vec{E}\cdot\text{d}\vec{s}(t)&=&c^2\int_{V(t)}\text{d}^3\vec{r}\div{\bigg\{\curl\bigg[\vec{B}-\frac{\vec{v}'\times\vec{E}}{c^2}\bigg]\bigg\}}\nonumber\\&&+\frac{1}{\epsilon_0}\oint_{s(t)}\text{d}\vec{s}(t)\cdot\big[\rho\vec{v}'-\vec{J}\big].
\end{eqnarray}
Interestingly, the 1st term on the right-hand side of Eqn.~(\ref{eqn8}) is zero as the divergence of the curl of a vector field is identically zero. Hence, the deformation of the moving Gaussian surface has no effect on the time-dependence of Gauss' flux integral. Now, it is to be noted that $\vec{J}(\vec{r},t)-\rho\vec{v}'(\vec{r},t)$ is the net outward current density with respect to the moving surface at the point $\vec{r}=\vec{r}'$ at  time $t$.  Thus, the integral on the 2nd term on the right-hand side of Eqn.~(\ref{eqn8}) is nothing but the net current ($I^{(s)}_{\text{in}}(t)=\oint_{s(t)}\text{d}\vec{s}(t)\cdot\big[\rho\vec{v}'-\vec{J}]$) flowing into the entire moving surface at time $t$.  Thus, we recast Eqn.~(\ref{eqn8}) as
\begin{eqnarray}\label{eqn9}
\frac{\text{d}}{\text{d}\text{t}}\oint_{s(t)}\vec{E}\cdot\text{d}\vec{s}(t)=\frac{I^{(s)}_{\text{in}}(t)}{\epsilon_0}.
\end{eqnarray}
This is the evolution equation for time-dependent Gauss' flux-integral. Now, integrating both sides of Eqn.~(\ref{eqn9}) with respect to time, we get 
\begin{eqnarray}\label{eqn10}
\oint_{s(t)}\vec{E}\cdot\text{d}\vec{s}(t)-\oint_{s(t_0)}\vec{E}\cdot\text{d}\vec{s}(t_0)=\frac{\triangle q_{\text{in}}(t,t_0)}{\epsilon_0}
\end{eqnarray}
where $\triangle q_{\text{in}}^{(s)}(t,t_0)=\int_{t_0}^tI^{(s)}_{\text{in}}(t')\text{d}t'$ is the net charge accumulated inside the Gaussian surface $s(t)$ through the entire moving surfaces $\{s(t')\}$ in the course of time ($t'$) for $t_0< t'<t$. Let us now suitably choose the reference time $t_0$ such that both the charges and the Gaussian surface were static from $t\rightarrow-\infty$ to $t=t_0$ with the total amount of charge $q_{in}^{(0)}$ in it. Thus, the flux-integral $\oint_{s(t_0)}\vec{E}\cdot\text{d}\vec{s}(t_0)$ follows Gauss' law $\oint_{s(t_0)}\vec{E}\cdot\text{d}\vec{s}(t_0)=\frac{q_{\text{in}}^{(0)}}{\epsilon_0}$ as mentioned in Eqn. (\ref{eqn5}) for electrostatics \cite{Jackson,Griffiths}. Hence, we recast Eqn. (\ref{eqn10}) as 
\begin{eqnarray}\label{eqn11}
\oint_{s(t)}\vec{E}(\vec{r},t)\cdot\text{d}\vec{s}(t)=\frac{q_{\text{in}}(t)}{\epsilon_0}
\end{eqnarray}
where $q_{\text{in}}(t)=q_{\text{in}}^{(0)}+\triangle q_{\text{in}}^{(s)}(t,t_0)$ is the net charge inside the Gaussian surface $s(t)$ at any instant of time $t$ for $t>t_0$. Eqn.~(\ref{eqn11}) is the desired result for a generic non-static case of charges moving inside and outside an arbitrarily moving (expanding/contracting and deforming) Gaussian surface. By comparing Eqns.~(\ref{eqn11}) and (\ref{eqn5}), we can safely say that the original form of Gauss' law remains unaltered even in a generic non-static case. The total amount of charges inside the moving surface $s(t')$ has not necessarily been considered to be conserved for the derivation of Eqn.~(\ref{eqn11}). It would have been conserved if we had considered $I^{(s)}_{\text{in}}(t')=0$ for $t_0<t'<t$. 

Let us now consider a special case of the conservation of the total charge inside the moving Gaussian surface. It is interesting to note from Eqn.~(\ref{eqn11}) that, if the charges arbitrarily move inside and outside the arbitrarily moving surface $s(t')$ and no current passes through it (i.e. $I^{(s)}_{\text{in}}(t')=0$) for $t_0<t'<t$,  then we have $\triangle q_{\text{in}}^{(s)}(t,t_0)=0$ for $t_0<t'<t$. In such a special case, Eqn.~(\ref{eqn11}) takes the form
\begin{eqnarray}\label{eqn12}
\oint_{s(t)}\vec{E}\cdot\text{d}\vec{s}(t)=\frac{q^{(0)}_{\text{in}}}{\epsilon_0}.
\end{eqnarray}
It should be mentioned that the flux of the electric field through a part of the moving Gaussian surface $s(t')$ though changes in time for $t_0<t'<t$, the flux through the whole Gaussian surface, of course, does not change at all in such a special case. Eqn.~(\ref{eqn12}) is compatible with the previous study for the even more special cases of the uniform (relativistic) motion of a charge inside a static Gaussian surface and the accelerated motion of a radiating charge inside a static Gaussian surface \cite{Garcia}. 

The time-dependent form of Gauss' flux-integral (Gauss' law), i.e. Eqn.~(\ref{eqn11}), obtained by us  although distinguished with the form $q_{\text{in}}(t)=q_{\text{in}}^{(0)}+\triangle q_{\text{in}}^{(s)}(t,t_0)$, is essentially not a new result because it exactly matches with Eqn.~(\ref{eqn5a}) for the form $q_{\text{in}}(t)=\int_{V(t)}\rho(\vec{r},t)\text{d}^3\vec{r}$. Thus, a question may arise whether we can derive Eqn.~(\ref{eqn9}) directly from Eqn.~(\ref{eqn5a}). Time-derivative of both sides of Eqn.~(\ref{eqn5a}) and evaluation of $\frac{\text{d}q_{\text{in}}(t)}{\text{d}t}=\frac{\text{d}}{\text{d}t}\int_{V(t)}\rho(\vec{r},t)\text{d}^3\vec{r}=\int_{V(t)}[\frac{\partial \rho}{\partial t}+\vec{v}'\cdot\grad{\rho}]\text{d}^3\vec{r}$ with uses of the continuity equation $\frac{\partial\rho}{\partial t}+\grad\cdot\vec{J}=0$ (or else Eqns.~(\ref{eqn1}) and (\ref{eqn4})), $\vec{v}'\cdot\grad\rho=\grad\cdot(\rho\vec{v}')$, and Gauss' divergence theorem, result in $\frac{\text{d}}{\text{d}\text{t}}\oint_{s(t)}\vec{E}\cdot\text{d}\vec{s}(t)=\frac{1}{\epsilon_0}\oint_{s(t)}\text{d}\vec{s}(t)\cdot\big[\rho\vec{v}'-\vec{J}\big]=\frac{I^{(s)}_{\text{in}}(t)}{\epsilon_0}$ which, as described below Eqn.~(\ref{eqn9}), is nothing but Eqn.~(\ref{eqn11}). It should be mentioned that the result,  $\frac{\text{d}}{\text{d}t}\int_{V(t)}\rho(\vec{r},t)\text{d}^3\vec{r}=\oint_{s(t)}\text{d}\vec{s}(t)\cdot\big[\rho\vec{v}'-\vec{J}\big]$, was previously obtained in Ref.\cite{Rothwell} without an explicit derivation. This result, however, is compatible with the Reynolds transport theorem which is usually applied in the field of fluid mechanics \cite{Reynolds}.

\begin{figure}
\includegraphics[width=8.5cm,height=7.0cm]{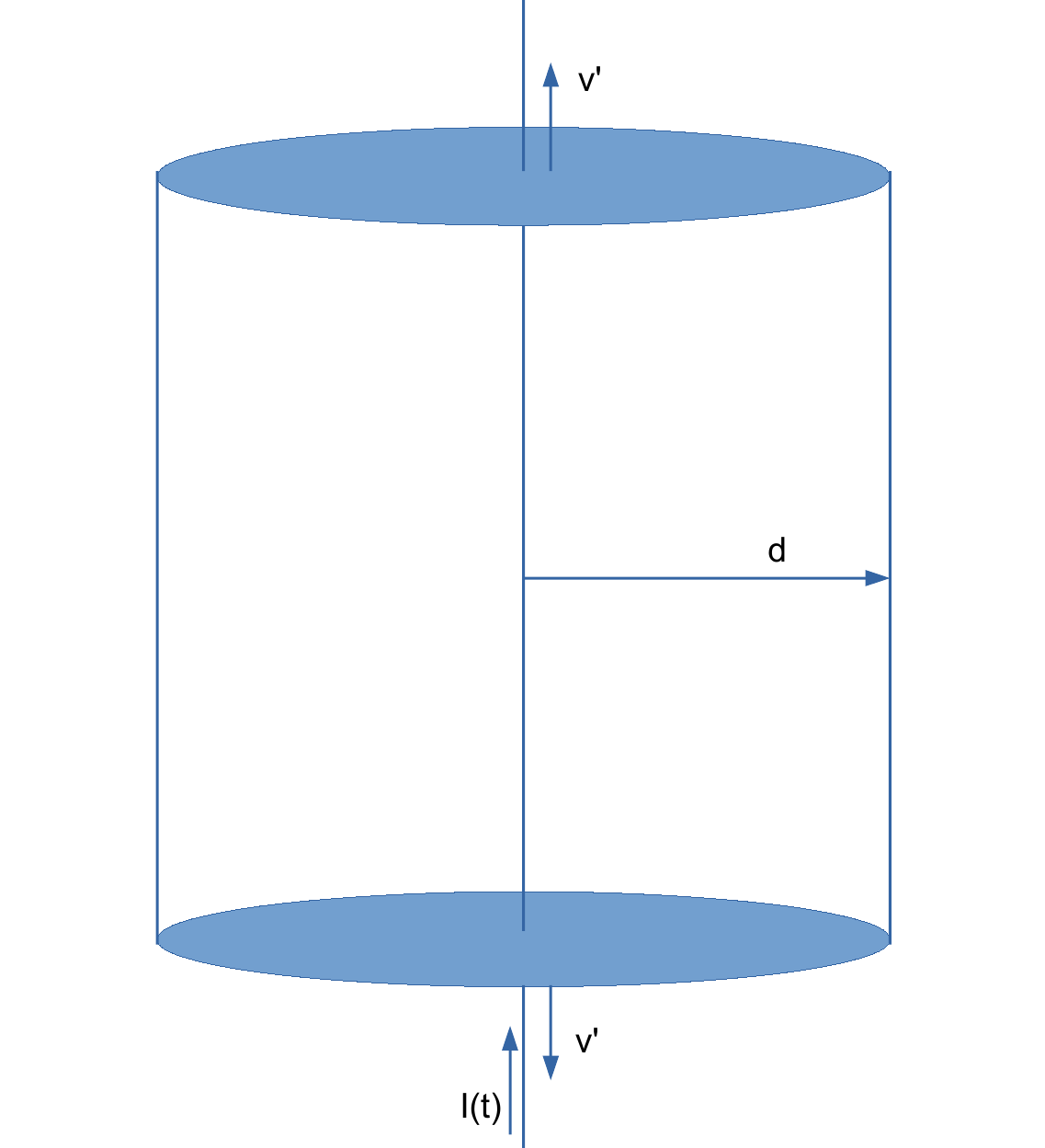}
\caption{An example with a cylindrical shaped Gaussian surface expanding along the flat surfaces. 
}
\label{fig2}
\end{figure}

To add a few simple illustrative examples in support of Eqn.~(\ref{eqn11}), we can consider a cylindrical-shaped Gaussian surface expanding along the flat surfaces with a constant speed $v'$ as depicted in figure \ref{fig2}. An infinitely long (1-D) straight electrical wire carrying current $I(t)$ is passing through the $z$-axis of the Gaussian surface. Let $d$ be the radius of each of the flat surfaces. Let us also take $I(t)=0$ for $t\le0$ and $I(t)=I_0$ (constant) for $t>0$. The electric field, being a physical quantity, is independent of the motion of the Gaussian surface because it is an imaginary surface. The electric field at a distance $d$ from the wire is obtained in Ref.\cite{Griffiths3} as $\vec{E}(\vec{r},t)|_{r=d}=-\frac{\mu_0I_0c}{2\pi\sqrt{[ct]^2-d^2}}\hat{z}$ for $t>d/c$. Incidentally, the electric field is perpendicular to the curved surface. The fluxes of the electric field passing through the two flat surfaces are also cancelling each other. Thus, we have $\oint_{s(t)}\vec{E}\cdot\text{d}\vec{s}(t)=0$ as expected because the total charge $q_{\text{in}}(t)$ (as well as the total charge density $\rho(\vec{r},t)$) inside the wire is zero all the time. 

We can further consider an expanding spherical Gaussian surface $s(t)$ enclosing a stationary charge $q$ at its center $\vec{r}=0$. Like the above, the electric field $\vec{E}(\vec{r},t)=\frac{q\hat{r}}{4\pi\epsilon_0r^2}$ remains independent of time in this case too. Hence, the flux integral $\oint_{s(t)}\vec{E}(\vec{r},t)\cdot\text{d}\vec{s}(t)=\frac{q}{\epsilon_0}$ remains the same all the time as the stationary charge never leaves the expanding Gaussian surface. If the stationary charge $q$ initially were kept outside the expanding Gaussian surface, it would eventually come inside the Gaussian surface, resulting in a change in the flux integral from $0$ to $q/\epsilon_0$. Similarly, the contraction of the Gaussian surface may also change the flux integral. The deformation of a Gaussian surface, on the other hand, doesn't affect the flux integral because the total charge inside the Gaussian surface, like the total mass of a rubber band/block, remains unaltered under any deformation.

\section{Conclusions}
To conclude, we have provided a detailed and pedagogically motivated re-examination of the integral form of Gauss’ law in situations where charges are moving arbitrarily inside and outside of an arbitrarily moving Gaussian surface which may expand/contract and deform. We have explicitly determined the time-dependence of Gauss' flux integral for such a generic non-static case. We have demonstrated the same with a few simple illustrative examples. Gauss' law, even in the generic non-static case, retains exactly the same structure as in the static case.

Our calculations are based on the Maxwell equations which don't need any relativistic correction. Hence, the final result (i.e. Eqn.~(\ref{eqn11})) on the time-dependent form of Gauss' law is essentially correct from the relativistic point of view, as an electric charge is a Lorentz invariant quantity.

Since the Gaussian surface is an imaginary surface, its motion does not affect the electric field. The flux integral of the same electric field, however, changes with time whenever there is a change in the total amount of stationary/moving charges inside the moving Gaussian surface.

While the uniformly moving charges result in no radiation, the accelerating charges inside and outside the Gaussian surface result in electromagnetic radiation. This radiation does not, however, change the form of Gauss' law as clear from Eqn.~(\ref{eqn11}).

Finally, it should be mentioned that the evolution equation (i.e. Eqn.~(\ref{eqn9})) obtained by us is novel, elegant, and important for determining the time-dependence of Gauss' flux integral in a generic non-static case.  Here, the use of the convective derivative $\frac{D\rho}{D t}=\frac{\partial \rho}{\partial t}+\vec{v}'\cdot\grad{\rho}$ \cite{Jackson1} of the scalar field $\rho(\vec{r},t)$ captures the effect of the expansion/contraction, not the deformation. Here, we haven't separately considered any overall displacement of the Gaussian surface because the expansion/contraction and the deformation together may result in an overall displacement. However, why there is no effect of the deformation on the flux integrals (Eqns.~(\ref{eqn9}) and (\ref{eqn11})), can be explained within our method with Eqn.~(\ref{eqn6}) as a starting point under the consideration of the convective derivative of the vector field $\vec{E}(\vec{r},t)$. Thus, our explicit derivation of the flux-integral for the moving surface can be directly useful in an undergraduate (as well as postgraduate) physics class for a better understanding of Gauss' law for electrodynamics. 

\section*{Acknowledgement}
S. Biswas acknowledges the partial financial support of the SERB (now ANRF), DST, Govt. of India under the EMEQ Scheme [No. EEQ/2023/000788]. We thank Mr. Bikram Keshari Behera (UoH, India) for helping us draw figure \ref{fig1}. Useful discussions with Dr. Abhiram Soori (UoH, India), Dr. Sudipto Muhuri (UoH, India), Dr. N. Sri Ram Gopal (UoH, India), Prof. V. Subrahmanyam (UoH, India), Mr. Rhitabrata Bhattacharyya (KCDCC(P), WB, India), and Prof. Koushik Ray (IACS, India) are gratefully acknowledged. We thank the anonymous reviewers for their thorough review and highly appreciate their critical remarks which have significantly contributed to improving the quality of the article.

\section*{Data availability statement}
No new data were created or analysed in this study.

%\section*{Author Declarations}
%\textbf{Conflict of Interest}: The author has no conflicts to disclose.

\end{document}